\documentclass[a4paper,10pt]{article}
\usepackage[dvips]{graphicx}
\usepackage{amssymb,amsmath}
\numberwithin{equation}{section}

\begin{document}

\title{ Accelerated expansion of the Universe driven by   G-essence}
\author{ K. K. Yerzhanov, P. Yu. Tsyba,  Sh. R. Myrzakul, I. I. Kulnazarov, R. Myrzakulov\footnote{The corresponding author. Email: rmyrzakulov@csufresno.edu; cnlpmyra1954@yahoo.com}\\ \textit{Eurasian International Center for Theoretical Physics, Dep. Gen. $\&$ Theor.  } \\ \textit{ Phys., Eurasian National University, Astana 010008, Kazakhstan} }

\date{}

\maketitle
\begin{abstract}In the present work  we analyze  the g-essence model for the particular Lagrangian: $L=R+2[\alpha X^n+\epsilon Y-V(\psi,\bar{\psi})]$. The g-essence models were proposed recently as an alternative and as a generalization to the scalar k-essence.   We have presented the 3 types solutions of the g-essence model. 
We reconstructed the corresponding potentials and the dynamics of the scalar and fermionic fields according the evolution of the scale factor.  The obtained results shows that the g-essence  model can describes the decelerated and accelerated expansion phases of the universe. 

\end{abstract}
\vspace{2cm} 

\sloppy

\tableofcontents
\section{Introduction} 
More than ten years after its initial discovery \cite{Perlmutter}-\cite{Riess}, cosmic acceleration remains an unsolved problem. In fact, this phenomenon is so much at odds with conventional particle physics and cosmology that a solution might require  a complete reformulation of the laws of physics governing both very small scales and very large scales. The contemporary models trying to explain cosmic acceleration using quantum field theory and general relativity fail to provide a convincing framework. The observational evidence from different sources for the present stage of accelerated expansion of our universe has driven the quest for theoretical explanations of such feature.  At present, theoretical physics are faced with two severe theoretical difficulties, that can be summarized as the dark energy and the dark matter problems.  Several theoretical models, responsible for this accelerated expansion, have been proposed in the literature, in particular, models  with some sourses and modified gravity, amongst others.  The simplest model of dark energy is the cosmological constant, which is a key ingredient in the $\Lambda$CDM model. Although the $\Lambda$CDM model is consistent very well with all observational data, it faces with the fine tuning problem.

During last years theories described by the action with the non-canonical  kinetic terms, k-essence, attracted a considerable interest.  Such theories were first studied in the context of k-inflation \cite{Mukhanov1}, and then the k-essence models were suggested  as dynamical dark energy for solving the cosmic coincidence problem \cite{Mukhanov2}-\cite{Chiba}.  

 In the recent years several approaches were made to explain the accelerated expansion by choosing fermionic fields as the gravitational sources of energy (see e.g. refs. \cite{Ribas1}-\cite{Armendariz-Picon}). In particular, it was shown that the fermionic  field plays very important role in: i) isotropization of initially anisotropic spacetime; ii) formation of singularity free cosmological solutions; iii) explaining late-time acceleration.  
 Quite recently,  the fermionic counterpart of the scalar k-essence was presented in \cite{MR} and called for short \textit{f-essence}. A  dark energy model, so-called \textit{g-essence}, has been proposed in \cite{MR} which is   the more general essence model.  In the present work, we construct   the some  cosmological solutions of the g-essence for the Lagrangian: $L=R+2[\alpha X^n+\epsilon Y-V(\psi,\bar{\psi})]$.     
The  formulation of the gravity-fermionic theory has been discussed in detail elsewhere \cite{Weinberg}-\cite{Birrell}., so we will only present the result here.
 
\section{G-essence}

Let us consider  the  M$_{34}$ - model. It  has  the  following action \cite{MR}
\begin {equation}
S=\int d^{4}x\sqrt{-g}[R+2K(X, Y, \phi, \psi, \bar{\psi})],
\end{equation} 
 where $K$ is some function of its arguments, $\phi$ is a scalar function, $\psi=(\psi_1, \psi_2, \psi_3, \psi_4)^{T}$  is a fermionic function  and $\bar{\psi}=\psi^+\gamma^0$ is its adjoint function. Here
\begin {equation}
X=0.5g^{\mu\nu}\nabla_{\mu}\phi\nabla_{\nu}\phi,\quad Y=0.5i[\bar{\psi}\Gamma^{\mu}D_{\mu}\psi-(D_{\mu}\bar{\psi})\Gamma^{\mu}\psi]
\end{equation}
are  the canonical kinetic terms for the scalar and fermionic fields, respectively. $\nabla_{\mu}$ and  $ D_{\mu}$ are the covariant derivatives. The model (2.1) admits important two reductions: \textit{k-essence} and \textit{f-essence} (see below). In this sense,  it is the more general essence model and in \cite{MR} it was   called \textit{g-essence}. 

The variation of the action (2.1) with respect to $g_{\mu\nu}$ gives us the following energy-momentum tensor for the g-essence fields:
$$
T_{\mu\nu}\equiv\frac{2}{\sqrt{-g}}\frac{\delta S}{\delta g_{\mu\nu}}=K_{X}\nabla_{\mu}\phi\nabla_{\nu}\phi+0.5iK_{Y}\left[\bar{\psi}\Gamma_{(\mu}D_{\nu)}\psi-D_{(\mu}\bar{\psi}\Gamma_{\nu)}\psi\right]
$$
	\begin{equation}
	-g_{\mu\nu}K=2K_{X}Xu_{1\mu}u_{1\nu}+K_{Y}Yu_{2\mu}u_{2\nu}-Kg_{\mu\nu},
\end{equation}
where $K_{X}={\partial K}/{\partial X}, K_{Y}={\partial K}/{\partial Y}, u_{1\mu}=\nabla_{\mu}\phi/\sqrt{2X}$ etc.  
The equation of motion for the scalar field $\phi$ is obtained by variation of the action (2.1) with respect to $\phi$,
	\begin{equation}-\frac{1}{\sqrt{-g}}\frac{\delta S}{\delta \phi}=(K_{X}g^{\mu\nu}+K_{XX}\nabla^{\mu}\phi\nabla^{\nu}\phi)\nabla_{\mu}\nabla_{\nu}\phi+2XK_{X\phi}-K_{\phi}.
\end{equation} Varying the action (2.1) with respect to $g_{\mu\nu}$ we get  the Einstein equations
\begin{equation}-\frac{2}{\sqrt{-g}}\frac{\delta S}{\delta g^{\mu\nu}}=R_{\mu\nu}-0.5Rg_{\mu\nu}-T_{\mu\nu}=0,
\end{equation}
where $R_{\mu\nu}$ is the Ricci tensor. Similarly, from the Euler-Lagrange equations applied to the Lagrangian density $K$ we can obtain the Dirac equations for the fermionic field $\psi$ and its adjoint $\bar{\psi}$ coupled to the gravitational and scalar  fields.

With the general formalism described above, we are now interested to investigate cosmology. We now  consider the dynamics of the  homogeneous, isotropic and flat FRW universe filled with g-essence. In this case, the background line element reads
	\begin{equation}
ds^2=dt^2-a^2(dx^2+dy^2+dz^2)
\end{equation}
and the vierbein is chosen to be
	\begin{equation}
	(e_a^\mu)=diag(1,1/a,1/a,1/a),\quad 
(e^a_\mu)=diag(1,a,a,a).
\end{equation}

 In the case of the FRW metric (2.6), the equations corresponding to the action (2.1) look  like \cite{MR}
	\begin{eqnarray}
	3H^2-\rho&=&0,\\ 
		2\dot{H}+3H^2+p&=&0,\\
		K_{X}\ddot{\phi}+(\dot{K}_{X}+3HK_{X})\dot{\phi}-K_{\phi}&=&0,\\
		K_{Y}\dot{\psi}+0.5(3HK_{Y}+\dot{K}_{Y})\psi-i\gamma^0K_{\bar{\psi}}&=&0,\\ 
K_{Y}\dot{\bar{\psi}}+0.5(3HK_{Y}+\dot{K}_{Y})\bar{\psi}+iK_{\psi}\gamma^{0}&=&0,\\
	\dot{\rho}+3H(\rho+p)&=&0,
	\end{eqnarray} 
where  the kinetic terms, the energy density  and  the pressure  take the form
\begin{equation}
X=0.5\dot{\phi}^2,\quad  Y=0.5i(\bar{\psi}\gamma^{0}\dot{\psi}-\dot{\bar{\psi}}\gamma^{0}\psi)
  \end{equation}
  and 
\begin{equation}
\rho=2K_{X}X+K_{Y}Y-K,\quad
p=K.
\end{equation}

 Note that the equations of the M$_{34}$ - model (2.8)-(2.13) can be rewritten as	
	\begin{eqnarray}
	3H^2-\rho&=&0,\\ 
		2\dot{H}+3H^2+p&=&0,\\
		(a^3K_{X}\dot{\phi})_{t}-a^3K_{\phi}&=&0,\\
		(a^3K_{Y}\psi^2_j)_{t}-2iK_{\bar{\psi}}(\gamma^0\psi)_j&=&0,\\ 
(a^3K_{Y}\psi^{*2}_j)_{t}+2iK_{\psi}(\bar{\psi}\gamma^{0})_j&=&0,\\
	\dot{\rho}+3H(\rho+p)&=&0.
	\end{eqnarray} Finally we present the following useful formula
	\begin{equation}
K_{Y}Y=0.5iK_{Y}(\bar{\psi}\gamma^{0}\dot{\psi}-\dot{\bar{\psi}}\gamma^{0}\psi)=-0.5(K_{\psi}\psi+K_{\bar{\psi}}\bar{\psi})
  \end{equation}
  and the equation for $u=\bar{\psi}\psi$:
  \begin{equation}
[\ln{(ua^3K_Y)}]_tu=-iK^{-1}_{Y}(\bar{\psi}\gamma^{0}K_{\bar{\psi}}-K_{\psi}\gamma^0\psi).
  \end{equation}
\subsection{Purely kinetic g-essence}
Let us consider the purely kinetic case of the M$_{34}$ - model that is when $K=K(X,Y)$. In this case, the system (2.8)-(2.13) becomes		
	\begin{eqnarray}
	3H^2-\rho&=&0,\\ 
		2\dot{H}+3H^2+p&=&0,\\
		a^3K_{X}\dot{\phi}-\sigma &=&0,\\
		a^3K_{Y}\psi^2_j-\varsigma_j &=&0,\\ 
a^3K_{Y}\psi^{*2}_j-\varsigma^{*}_j&=&0,\\
	\dot{\rho}+3H(\rho+p)&=&0,
	\end{eqnarray}
	where $\sigma$  $(\varsigma)$ is the real (complex) constant. Hence we immediately get the solutions of the  Klein-Gordon and Dirac equations, respectively, as
	\begin{equation}
\phi=\sigma\int\frac{dt}{ a^3K_{X}},\quad \psi_j=\sqrt{\frac{\varsigma_j}{a^3K_{Y}}}.
\end{equation}
 Also the following useful formula takes place
 	\begin{equation}
X=\frac{0.5\sigma^2}{a^6K_{X}^2} \quad or \quad K_X=\frac{\sigma}{a^3\sqrt{2X}}.
\end{equation}
It is interesting to note that for the purely kinetic g-essence the solutions of the  Klein-Gordon and Dirac equations are related by the formula
	\begin{equation}
\dot{\phi}=\sigma\varsigma_j^{-1}\psi_{j}^2.
\end{equation}
Let us  conclude this section as: for the purely kinetic case $K=K(X,Y)$ from (2.22) follows  that $Y=0$ so that in fact we have $K=K(X,Y)=K(X, 0)=K(X)$. So we will go further, having passed by this case.

\subsection{K-essence}

Let us now we consider the following particular case of the M$_{34}$ - model (2.1):
\begin{equation}
K=K_1=K_1(X,\phi)
\end{equation} that corresponds to k-essence.
Then the system (2.8)-(2.13) takes the form of the equations of   k-essence (see e.g.  \cite{Mukhanov1}-\cite{Chiba})
\begin{eqnarray}
	3H^2-\rho_k &=&0,\\ 
		2\dot{H}+3H^2+p_k&=&0,\\
		K_{1X}\ddot{\phi}+(\dot{K}_{1X}+3HK_{1X})\dot{\phi}-K_{1\phi}&=&0,\\
	\dot{\rho}_k+3H(\rho_k+p_k)&=&0,
	\end{eqnarray} 
where  the energy density  and  the pressure  are given by
\begin{equation}
\rho_k=2K_{1X}X-K_1,\quad
p_k=K_1.
\end{equation} 
As is well-known, the  energy-momentum tensor for the k-essence field has the form
	\begin{equation}
T_{\mu\nu}=K_{X}\nabla_{\mu}\phi\nabla_{\nu}\phi
	-g_{\mu\nu}K=2K_{X}Xu_{1\mu}u_{1\nu}-Kg_{\mu\nu}=(\rho_k+p_k)u_{1\mu}u_{1\nu}-p_kg_{\mu\nu}.
\end{equation}

It is interesting to note that in the case of  the FRW metric (2.6), purely kinetic k-essence and F(T) - gravity (modified teleparallel gravity) are eqivalent  to each other, if $a=e^{\pm\frac{\phi-\phi_0}{\sqrt{12}}}$ \cite{MR1}-\cite{MR2}.

\subsection{F-essence}

Now we consider the following reduction of the M$_{34}$ - model (2.1):
\begin{equation}
K=K_2=K_2(Y,\psi, \bar{\psi})
\end{equation}
that corresponds to the M$_{33}$ - model that is  the f-essence \cite{MR}. The  energy-momentum tensor for the f-essence field has the form
$$
T_{\mu\nu}\equiv\frac{2}{\sqrt{-g}}\frac{\delta S}{\delta g_{\mu\nu}}=0.5iK_{Y}\left[\bar{\psi}\Gamma_{(\mu}D_{\nu)}\psi-D_{(\mu}\bar{\psi}\Gamma_{\nu)}\psi\right]-
$$
	\begin{equation}	-g_{\mu\nu}K=K_{Y}Yu_{2\mu}u_{2\nu}-Kg_{\mu\nu}=(\rho_f+p_f)u_{2\mu}u_{2\nu}-p_fg_{\mu\nu}.
\end{equation}
For the FRW metric (2.6),  the equations of the f-essence  become \cite{MR}
\begin{eqnarray}
	3H^2-\rho_f &=&0,\\ 
		2\dot{H}+3H^2+p_f&=&0,\\
		K_{2Y}\dot{\psi}+0.5(3HK_{2Y}+\dot{K}_{2Y})\psi-i\gamma^0K_{2\bar{\psi}}&=&0,\\ 
K_{2Y}\dot{\bar{\psi}}+0.5(3HK_{2Y}+\dot{K}_{2Y})\bar{\psi}+iK_{2\psi}\gamma^{0}&=&0,\\
	\dot{\rho}_f+3H(\rho_f+p_f)&=&0,
	\end{eqnarray} 
where 
\begin{equation}
\rho_f=K_{2Y}Y-K_2,\quad
p_f=K_2.
\end{equation}

\section{Solutions} Let us we present some solution of the g-essence (2.1). To do it, we consider the case
	\begin{equation}
K=K(X,Y, \psi, \bar{\psi})=\alpha X^n+\epsilon Y-V(\psi, \bar{\psi}).
\end{equation}
Then the system (2.8)-(2.13) takes the form
\begin{eqnarray}
	3H^2-\rho&=&0,\\ 
		2\dot{H}+3H^2+p&=&0,\\
		\ddot{\phi}+[3H+(n-1)(\ln{X})_{t}]\dot{\phi}&=&0,\\
		\dot{\psi}+1.5H\psi+i\epsilon^{-1}\gamma^0V_{\bar{\psi}}&=&0,\\ 
\dot{\bar{\psi}}+1.5H\bar{\psi}-i\epsilon^{-1}V_{\psi}\gamma^{0}&=&0,\\
	\dot{\rho}+3H(\rho+p)&=&0,
	\end{eqnarray} 
	where
	\begin{equation}
\rho=\alpha(2n-1)X^n+V, \quad
p=\alpha X^n+\epsilon Y-V.
\end{equation}
It has the following solution
\begin{eqnarray}
	X&=&\sqrt[2n-1]{\frac{\sigma^2}{2n^2\alpha^2 a^6}},\\ 
	Y&=&-2\epsilon^{-1}[\dot{H}+\alpha n(\frac{\sigma^2}{2n^2\alpha^2a^6})^{\frac{n}{2n-1}}],\\ 
	V&=&3H^2-(2n-1)\alpha (\frac{\sigma^2}{2n^2\alpha^2 a^6})^{\frac{n}{2n-1}},\\
	K&=&-2\dot{H}-3H^2.
	\end{eqnarray}

Now we would like to present  some explicit solutions. Consider examples.

\subsection{Example 1: $a=a_0t^{\lambda}$} 	
Let us first consider the power-law solution
\begin{equation}
	a=a_0t^{\lambda}.
\end{equation} Then we get
\begin{eqnarray}
	X&=&\sqrt[2n-1]{\frac{\sigma^2}{2n^2\alpha^2 a_0^6t^{6\lambda}}},\\ 
	Y&=&-2\epsilon^{-1}[-\frac{\lambda}{t^2}+\alpha n(\frac{\sigma^2}{2n^2\alpha^2a_0^6t^{6\lambda}})^{\frac{n}{2n-1}}],\\ 
	V&=&\frac{3\lambda^2}{t^2}-(2n-1)\alpha (\frac{\sigma^2}{2n^2\alpha^2 a_0^6t^{6\lambda}})^{\frac{n}{2n-1}},\\
	K&=&\frac{\lambda(2-3\lambda)}{t^2}.
	\end{eqnarray}
	Let us simplify the problem assuming that the potential has the form
$V=V(u)$. Then
from (2.22)-(2.23) follows that 
\begin{equation}
	u=\frac{c}{\epsilon a^{3}}, \quad Y	=\epsilon^{-1}V_{u}u.
  \end{equation}
As 
\begin{equation}
	u=\frac{c}{\epsilon a_0^{3}t^{3\lambda}}, \quad 	t=[\frac{c}{\epsilon a_0^{3}u}]^{\frac{1}{3\lambda}}
  \end{equation}
  the expression for the potential takes the form
  \begin{equation}
	 V=3\lambda^2\left(\frac{\epsilon a_0^{3}u}{c}\right)^{\frac{2}{3\lambda}}-(2n-1)\alpha \left(\frac{\sigma^2\epsilon^2u^2}{2n^2\alpha^2 c^2}\right)^{\frac{n}{2n-1}}.
  \end{equation}
  So finally we get the following solutions of the gravitational, Klein-Gordon  and Dirac equations:
\begin{eqnarray}
a&=&a_0t^{\lambda},\\ 
\phi&=&\left(\frac{2n-1}{2n-1-3\lambda}\right)\sqrt[2(2n-1)]{\frac{2^{2(n-1)}\sigma^2}{n^2\alpha^2 a^6_0}}t^{\frac{2n-1-3\lambda}{2n-1}},\\ 
\psi_l&=&\frac{c_{l}}{a^{1.5}_0t^{1.5\lambda}}e^{-i\left[-\frac{2\lambda}{t}+\frac{\alpha(2n-1)}{3\lambda} \left(\frac{\sigma^2}{2n^2\alpha^2 a^6_0}\right)^{\frac{n}{2n-1}}t^{\frac{2n-1-6n\lambda}{2n-1}}
\right]}\quad (l=1,2),\\	
\psi_k&=&\frac{c_{k}}{a^{1.5}_0t^{1.5\lambda}}e^{i\left[-\frac{2\lambda}{t}+\frac{\alpha(2n-1)}{3\lambda} \left(\frac{\sigma^2}{2n^2\alpha^2 a^6_0}\right)^{\frac{n}{2n-1}}t^{\frac{2n-1-6n\lambda}{2n-1}}
\right]}\quad (k=3,4),
	\end{eqnarray}
where  $c_j$ obey the following condition
	\begin{equation}
c=|c_{1}|^2+|c_{2}|^2|-|c_{3}|^2-|c_{4}|^2.
\end{equation}
  If 
   \begin{equation}
	\lambda=\frac{2n-1}{3n},
  \end{equation}
  then
  \begin{eqnarray}
	X&=&\left(\sqrt[2n-1]{\frac{\sigma^2}{2n^2\alpha^2 a_0^6}}\right)t^{-\frac{2}{n}},\\ 
	Y&=&2\epsilon^{-1}\left[\frac{2n-1}{3n}-\alpha n\left(\frac{\sigma^2}{2n^2\alpha^2a_0^6}\right)^{\frac{n}{2n-1}}\right]t^{-2},\\ 
	V&=&(2n-1)\left[\frac{2n-1}{3n^2}-\alpha \left(\frac{\sigma^2}{2n^2\alpha^2 a_0^6}\right)^{\frac{n}{2n-1}}\right]t^{-2},\\
	K&=&\frac{2n-1}{3n^2}t^{-2},\\
	u&=&\frac{c}{\epsilon a_0^3}t^{\frac{1-2n}{n}}.
		\end{eqnarray}
		
		In this case the potential has the form
		\begin{equation}
	 V=(2n-1)\left[\frac{2n-1}{3n^2}-\alpha \left(\frac{\sigma^2}{2n^2\alpha^2 a_0^6}\right)^{\frac{n}{2n-1}}\right]\left(\frac{\epsilon a_0^3u}{c}\right)^{\frac{2n}{2n-1}}.
  \end{equation}
  
  Finally,   let us we present  the expressions for the equation of state and deceleration parameters. For the our particular solution  (3.13)  they take the form
	\begin {equation}
w=-1+\frac{2n}{2n-1}, \quad q=\frac{n+1}{2n-1}.\end{equation}
These formulas  tell us   that for $n\in (-1, 0.5) $ [$n\in (-\infty, -1)$ and $n\in (0.5, +\infty)$] we get the  accelerated [decelerated] expansion phase   of the universe.

\subsection{Example 2: $	a=a_0\sinh^{m}[\beta t]$}As the second example we consider the  solution
\begin{equation}
	a=a_0\sinh^{m}[\beta t].
\end{equation} 
In this case, we have
\begin{equation}
	H=m\beta \coth[\beta t], \quad \dot{H}=m\beta^2\sinh^{-2}[\beta t], \quad u=\frac{c}{\epsilon a_0^3\sinh^{3m}[\beta t]}
\end{equation} 
and
\begin{eqnarray}
	X&=&\sqrt[2n-1]{\frac{\sigma^2}{2n^2\alpha^2 a^6_0\sinh^{6m}[\beta t]}},\\ 
	Y&=&-2\epsilon^{-1}\left[m\beta^2\sinh^{-2}[\beta t]+\alpha n\left(\frac{\sigma^2}{2n^2\alpha^2a^6_0\sinh^{6m}[\beta t]}\right)^{\frac{n}{2n-1}}\right],\\ 
	V&=&3m^2\beta^2 \coth^{2}[\beta t]-(2n-1)\alpha \left(\frac{\sigma^2}{2n^2\alpha^2 a^6_0\sinh^{6m}[\beta t]}\right)^{\frac{n}{2n-1}},\\
	K&=&-2m\beta^2\sinh^{-2}[\beta t]-3m^2\beta^2 \coth^{2}[\beta t].
	\end{eqnarray}
So finally we get the following solutions of the g-essence:
\begin{eqnarray}
a&=&a_0\sinh^{m}[\beta t],\\ 
	\phi&=&\sqrt[2(2n-1)]{\frac{2^{2(n-1)}\sigma^2}{n^2\alpha^2 a^6_0}}\int\frac{dt}{\sinh^{\frac{3m}{2n-1}}[\beta t]},\\ 
	\psi_l&=&\frac{c_{l}}{a^{1.5}_0\sinh^{1.5m}[\beta t]}e^{-iD}\quad (l=1,2),\\	
	\psi_k&=&\frac{c_{k}}{a^{1.5}_0\sinh^{1.5m}[\beta t]}e^{iD}\quad (k=3,4)
	\end{eqnarray}
	and the following expression for the potential 
	\begin{equation}
	V=3m^2\beta^2 \left(1+\sqrt[3m]{\frac{\epsilon^2a_0^6u^2}{c^2}}\right)-\alpha(2n-1) \left(\frac{\epsilon^2\sigma^2u^2}{2n^2\alpha^2 c^2}\right)^{\frac{n}{2n-1}}.
\end{equation} Here 
	\begin{equation}
D=-\frac{2\epsilon a_0^3}{c}\int\left[m\beta^2\sinh^{3m-2}[\beta t]+\alpha n\left(\frac{\sigma^2}{2n^2\alpha^2 a^6_0}\right)^{\frac{n}{2n-1}}\sinh^{-\frac{3m}{2n-1}}[\beta t]\right]dt
\end{equation}
and $c_j$ obey the  condition (3.25). The expressions for the equation of state and deceleration parameters  take the form
	\begin {equation}
w=-1-\frac{2}{3m}\tanh^2[\beta t], \quad q=-\frac{m-1+\tanh^2[\beta t]}{m}.\end{equation}
These formulas  tell us   that this solution can describes  the  accelerated and decelerated expansion phases   of the universe.

\subsection{Example 3: $	a=a_0e^{\beta t}$} Finally,  we consider the following  solution for the scale factor:
\begin{equation}
	a=a_0e^{\beta t} \quad (\beta=const).
\end{equation} 
In this case, we have
\begin{equation}
	H=\beta, \quad \dot{H}=0, \quad u=\frac{c}{\epsilon a_0^3e^{3\beta t}}
\end{equation} 
and
\begin{eqnarray}
	X&=&\sqrt[2n-1]{\frac{\sigma^2}{2n^2\alpha^2 a^6_0e^{6\beta t}}},\\ 
	Y&=&-2\epsilon^{-1}\alpha n\left(\frac{\sigma^2}{2n^2\alpha^2a^6_0e^{6\beta t}}\right)^{\frac{n}{2n-1}},\\ 
	V&=&3\beta^2 -(2n-1)\alpha \left(\frac{\sigma^2}{2n^2\alpha^2 a^6_0e^{6\beta t}}\right)^{\frac{n}{2n-1}},\\
	K&=&-3\beta^2.
	\end{eqnarray}
So finally we get the following solutions of the g-essence:
\begin{eqnarray}
a&=&a_0e^{\beta t},\\ 
	\phi&=&\sqrt[2(2n-1)]{\frac{2^{2(n-1)}\sigma^2}{n^2\alpha^2 a^6_0}}\frac{(1-2n)e^{\frac{3\beta}{1-2n}t}}{3\beta},\\ 
	\psi_l&=&\frac{c_{l}}{a^{1.5}_0\sinh^{1.5m}[\beta t]}e^{-i\left[\frac{2\epsilon \alpha n^2a_0^3}{3\beta c}\left(\frac{\sigma^2}{2n^2\alpha^2 a^6_0}\right)^{\frac{n}{2n-1}}e^{-\frac{3\beta t}{2n-1}}\right]}\quad (l=1,2),\\	
	\psi_k&=&\frac{c_{k}}{a^{1.5}_0\sinh^{1.5m}[\beta t]}e^{i\left[\frac{2\epsilon \alpha n^2a_0^3}{3\beta c}\left(\frac{\sigma^2}{2n^2\alpha^2 a^6_0}\right)^{\frac{n}{2n-1}}e^{-\frac{3\beta t}{2n-1}}\right]}\quad (k=3,4)
	\end{eqnarray}
	and the following expression for the potential 
	\begin{equation}
	V=3\beta^2 -\alpha(2n-1) \left(\frac{\epsilon^2\sigma^2u^2}{2n^2\alpha^2 c^2}\right)^{\frac{n}{2n-1}}.
\end{equation} As is well-known that in  this case the  equation of state and deceleration parameters are:
	\begin {equation}
w=-1, \quad q=-1.\end{equation}

	\section{Conclusion}
	In this work we studied the g-essence model for the particular Lagrangian: $L=R+2[\alpha X^n+\epsilon Y-V(\psi,\bar{\psi})]$ which involves the scalar and fermionic fields. The g-essence models were proposed recently as an alternative and as a generalization to scalar k-essence.   We have presented the 3 types solutions of the g-essence model. 
We reconstructed the corresponding potentials and the dynamics of the scalar and fermionic fields according the evolution of the scale factor. We calculated the  equation of state and deceleration parameters for the presented solutions. The obtained results tell us that the model can describes the decelerated and accelerated expansion phases of the universe. We want, however, to conclude with more conservative viewpoint that further work is needed to understand whether g-essence can be relevant in realistic cosmology indeed.

\end{document}